\documentclass[aps,amsfonts,epsf]{revtex4}
\usepackage{amsfonts,mathtools}
\usepackage{graphics}
\usepackage{graphicx}
\usepackage{epsfig}

\newcommand\beq{\begin{equation}}
\newcommand\eeq{\end{equation}}
\DeclareMathOperator*{\Ket}{\raisebox{-2pt}{\mbox{\Large $\mathbf K$}}}
\newcommand{\cfsum}[2]{\Ket_{#1}^{#2}}
\newcommand{\D}{\displaystyle}

\def\gsim{ \lower .75ex \hbox{$\sim$} \llap{\raise .27ex \hbox{$>$}} }
\def\lsim{ \lower .75ex\hbox{$\sim$} \llap{\raise .27ex \hbox{$<$}} }

\def\prd{Phys.~Rev.~D }       

\begin{document}
\input epsf.tex

\title{Calculating the continued fraction coefficients of a sub-diagonal Pad\'e approximant at arbitrary order.}
\author{J\'er\^ome Carr\'e\footnote{email : carre@apc.univ-paris7.fr} \& Edward K. Porter\footnote{email : porter@apc.univ-paris7.fr}\\}
\vspace{1cm}
\affiliation{Fran\c{c}ois Arago Center, APC, UMR 7164, Universit\'e Paris 7 Denis Diderot,\\ 10, rue Alice Domon et L\'{e}onie Duquet, 75205 Paris Cedex 13, France}
\vspace{1cm}
\begin{abstract}
The inspiral of two compact objects in gravitational wave astronomy is described by a post-Newtonian expansion in powers of $(v/c)$.  In most cases, it is believed that the post-Newtonian expansion is asymptotically divergent.  A standard technique for accelerating the convergence of a power series is to re-sum the series by means of a rational polynomial called a Pad\'e approximation.  If we liken this approximation to a matrix, the best convergence is achieved by staying close to a diagonal Pad\'e approximation.  This broadly presents two subsets of the
approximation : a super-diagonal approximation $P^M_N$ and a sub-diagonal approximation $P_M^N$, where $M = N+\epsilon$, and $\epsilon$ takes the 
values of 0 or 1.  Left as rational polynomials, the coefficients in both the numerator and denominator need to be re-calculated as the order of the initial power series approximation is increased.  However, the sub-diagonal Pad\'e approximant is computationally advantageous as it can be expressed in terms of a 
Gauss-like continued fraction.   Once in this form, each coefficient in the continued fraction is uniquely determined at each order.  This means that
as we increase the order of approximation of the original power series, we now have only one new additional coefficient to calculate in the continued fraction.  While it
is possible to provide explicit expressions for the continued fraction coefficients, they rapidly become unwieldy at high orders of approximation.  It is also possible to numerically calculate the coefficients by means of ratios of Hankel determinants.  However, these determinants can be ill-conditioned and lead to numerical instabilities.  In this article, we present
a method for calculating the continued fraction coefficients at arbitrary orders of approximation.

\end{abstract}

\maketitle

\section{Introduction.}
Gravitational wave (GW) astronomy is one of the newest and most exiting fields in astronomy and astrophysics.  At present,  a number of the ground based interferometers, such as Advanced LIGO and Advanced Virgo, are being upgraded with the goal of being back online in 2015~\cite{ligo,virgo}.  There are also plans for a cryogenic detector in Japan called LCGT~\cite{lcgt} and a large subterranean detector in Europe called The Einstein Telescope~\cite{ET}.  In order to detect more massive sources of GWs, there are also plans for an ESA led spaced based GW mission called eLISA~\cite{eLISA}.    One of the most powerful sources of interest is the inspiral and merger of compact binary systems such as binaries of supermassive black holes from galaxy mergers, or the inspiral of stellar mass black holes and neutron stars.  In all of these cases, the rate of energy loss due to the emission of GWs is given by the energy balance equation
\begin{equation}
m\frac{dE(v)}{dt} = -F(v),
\end{equation}
where $m$ is the total mass of the binary system, $E(v)$ is the binding energy of the system, $v$ is the linear velocity of the system components and $F(v)$ is the 
GW flux.  Most of the search methods in the field are based on optimal Wiener or matched filtering, where phase matching between the model template and a possible source in the data is the highest priority.  As the phase of the GW is a function of both $E(v)$ and $F(v)$, in order to successfully search for these systems in a data set, we need to be able to successfully model the GW flux function.  However, as we do not have
an exact solution for the GW flux from a binary system, it is represented by a post-Newtonian expansion (i.e. an expansion in powers of $(v/c)$, where $c$ is the speed of light).  This series has the
form
\begin{equation}
F(v) = F_N\left[\sum_{i=0}^{n}A_i v^i + \ln(v) \sum_{i=6}^{n}B_i v^i\right],
\label{eq:pnflux}
\end{equation}
where $n$ has different values depending on the type of physical system in question.  The quantity $F_N= 32\eta^2 v^{10} /5$ is the Newtonian flux and $\eta = m_1 m_2 /m^2$ is the symmetric mass ratio.  As the post-Newtonian approximation is based on the limiting condition $v\ll c$,  it is thought that the above representation of the flux is asymptotically divergent as we approach the domain where $v\approx c$.  An improved model, based on Pad\'e approximation, was proposed where the flux is written in the form~\cite{DIS}
\begin{equation}
F_p(v) = F_N \left(1-v/v_{lr}\right)^{-1}\left[1+\ln(v/v_{lso})\sum_{i=6}^n l_i v^i\right] C_n\left[\sum_{i=0}^n a_i x^i\right],
\label{eq:flux}
\end{equation}
where the coefficients are defined by $l_i=l_i(A_i, B_i)$ and $a_i = a_i(A_i, B_i, v_{lso})$.  This flux representation was demonstrated to have better convergence properties than the PN flux for both spinning and non-spinning test mass systems~\cite{DIS, PS}, and also for non-spinning comparable mass binaries~\cite{Buonanno}.  In Equation~(\ref{eq:flux}), the quantities $(v_{lso},v_{lr})$ correspond to the orbital velocities at the last stable and unstable circular orbits.  

In approximation theory one can equate a Pad\'e approximation to a power series by
\begin{equation}
\sum_{i=0}^{n} a_i x^i=P^N_M \equiv \frac{\displaystyle \sum_{i=0}^N d_i x^i}{\displaystyle \sum_{i=0}^M f_i x^i},
\end{equation}
such that the $n+1$ coefficients on the left hand side are matched by the $N+M+1$ coefficients on the right hand side.  Theory suggests that the best convergence is obtained when the Pad\'e approximation is kept to near diagonal values, i.e. $N\approx M$.  This allows us to define two types of approximant : sub-diagonal Pad\'e approximants $(M\geq N)$ and super-diagnoal Pad\'e approximants $(N \geq M)$.  The advantage of defining the two types of approximants is that we sometimes obtain poles in the Pad\'e approximation.  However, a pole in say the sub-diagnoal approximant is usually not reflected in the super-diagonal approximant.  In the particular case where $M=N$ or $M=N+1$, we can write the Pad\'e approximant in a continued fraction form.  The advantage of this is in the computation of coefficients.  In the case of a super-diagonal approximant, each time we increase the order of the power series, we have to re-compute all $N+M+1$ coefficients in the Pad\'e approximation.  However, in the sub-diagonal case, where we have re-written the approximation as a continued fraction, at each new order in the power series, we only have to calculate the newest $n^{th}$ order coefficient, as all others stay constant.  We can now define the operator $C_n[...]$ in Equation~(\ref{eq:flux}) as the continued fraction representation of a sub-diagonal Pad\'e approximant, i.e.
\begin{equation}
C_n\left[\sum_{i=0}^n a_i x^i\right] = \frac{c_{0}}{\D1 + 
                        \frac{c_{1}x}{\D1 + 
                         \frac{c_{2}x}{\D
                          \frac{\ddots}{{\D 1 + 
                                c_{n}x}}}}},
\label{eq:confrac}
\end{equation}


The continued fraction used for the flux model corresponds to a sub-diagonal Pad\'e approximant $P^{N}_{M}$ with $M = N + \epsilon$ and $\epsilon$ having values of 0 or 1, i.e.
\begin{equation}
C_n\left[\sum_{i=0}^n a_i x^i\right] \approx P^N_M\equiv\frac{\displaystyle \sum_{i=0}^N d_i x^i}{\displaystyle \sum_{i=0}^M f_i x^i}.
\label{eq:confrac}
\end{equation}
While the coefficients in the continued fraction can be solved for in terms of explicit expressions using a computer algebra system such as MAPLE or Mathematica, it is
difficult to present the expressions beyond a certain order due to their length and complexity.  Our goal here is to present a fast numerical algorithm to calculate the
coefficients of a continued fraction at arbitrary order $n$.

\section{Theoretical Foundations}
In cases where a power series has slow convergence, a rational function approximant such as Pad\'e approximation, and by extension a continued fraction representation,
have been shown to increase the speed of convergence~\cite{Acton}.  In general, a power series converges within a tight radius of convergence, but only close to the point of expansion and in
between the first poles.  A power series can also be numerically unstable due to the fact that, as we increase the number of terms in the expansion, the magnitudes of the exponents of the monomials continue to grow.    A continued fraction, on the other hand, converges everywhere in the complex plane, except at the poles.  However, as the continued fraction approximation to a power series contains a monomial in each denominator, they are quite adept at capturing the poles~\cite{Wall}.

While continued fractions have a number of attractive properties, there are certain numerical issues in calculating the continued fraction coefficients, given a generating power series.  We will briefly outline this in the following example.   Firstly, by using the following nomenclature~\cite{LW}
\begin{equation}
\cfsum{i=1}{\infty}\frac{a_i}{b_i} = \frac{a_1}{\D b_1 + \frac{a_2}{\D b_2 + \frac{a_3}{\ddots}}},
\end{equation}
we can define a regular continued fraction (from here-on, {\emph {C-fraction}})~\cite{JT}
\begin{equation}
\cfsum{i=1}{\infty}\frac{a_i x^{\alpha(i)}}{1} ,
\end{equation}
where $\alpha(i)$ is a constant.  If we now take $\alpha(i)=1$, and equate the {\emph {C-fraction}} with a power series, we can write
\begin{equation}
1+\sum_{i=1}^{\infty}\,c_i x^i = 1 + \cfsum{i=1}{\infty}\frac{a_i x}{1}.
\end{equation}
In this case, the coefficients of the {\emph {C-fraction}}, $a_i$, are given by~\cite{JT}
\begin{eqnarray}
a_1 &=& H_{1}^{(1)}, \\ \nonumber\\
a_{2m} &=& -\frac{H^{(1)}_{m-1}H^{(2)}_{m}}{H^{(1)}_{m}H^{(2)}_{m-1}} \,\,\,\,\,\,\, , \,\,\,\,\,\,\,\,a_{2m+1} = -\frac{H^{(1)}_{m+1}H^{(2)}_{m-1}}{H^{(1)}_{m}H^{(2)}_{m}},
\end{eqnarray}
where we define the Hankel determinant
\begin{equation}
H_m^{(i)} =H_m^{(i)} (c_i)\equiv det\begin{pmatrix} c_n & c_{n+1} & \cdots & c_{m+n-2} & c_{m+n-1} \\c_{n+1} & c_{n+2} & \cdots & c_{m+n-1} & c_{m+n} \\ \vdots & \vdots & \ddots & \vdots & \vdots \\  c_{m+n-1} & c_{m+n} & \cdots & c_{2m+n-3} & c_{2m+n-2}\end{pmatrix}.
\end{equation}
The problem with this method of finding the {\emph {C-fraction}} coefficients is that the Hankel determinants are frequently ill-conditioned, leading to a number of numerical issues.  We can see from the above expression that in the case of one of the Hankel determinants in the denominator being very small, the {\emph {C-fraction}} coefficient blows up.

\section{Calculating the rational function coefficients.}
Our goal in this work is to present a method that avoids using ratios of determinants and will provide a more numerically stable method of calculating the {\emph {C-fraction}} coefficients.  The first step in the analysis is to have expressions for the coefficients of the numerator and denominator in the rational polynomial defined by Equation~(\ref{eq:confrac}) at arbitrary order.  By expanding 
at lower orders of $n$, we find that the first five rational function expansions have the form
\begin{eqnarray}
P^0_1 &=& \frac{c_0}{1+c_1 x}, \nonumber\\
P^1_1 &=& \frac{c0\left(1+c_2 x\right)}{1+\left(c_1 + c_2\right)x},\nonumber\\
P^1_2 &=& \frac{c0\left(1+\left(c_2 + c_3\right) x\right)}{1+\left(c_1 + c_2 + c_3\right)x + c_1 c_3 x^2},\\
P^2_2 &=& \frac{c0\left(1+\left(c_2 + c_3 + c_4\right) x + c_2 c_4 x^2\right)}{1+\left(c_1 + c_2 + c_3 + c_4\right)x + \left(c_1 \left(c_3 + c_4\right) + c_2 c_4 \right)x^2},\nonumber\\
P^2_3 &=& \frac{c0\left(1+\left(c_2 + c_3 + c_4 + c_5\right) x + \left(c_2 \left(c_4 + c_5\right) + c_3 c_5\right)x^2\right)}{1+\left(c_1 + c_2 + c_3 + c_4 + c_5\right)x + \left(c_1 \left(c_3 + c_4 + c_5\right) + c_2 \left(c_4 + c_5\right) + c_3 c_5\right)x^2 + c_1 c_3 c_5 x^3}.\nonumber
\end{eqnarray}
By investigation, the $d_i$ coefficients in the numerator follow the form
\begin{eqnarray}
d_0 &=& c_0,\nonumber\\
d_1 &=& c_0 \sum_{j=2}^{N+M} c_j = c_0 \sum_{j=2}^n c_j,\nonumber\\
d_2 &=& c_0 \sum_{j=2}^{n-2} \sum_{k=j+2}^{n} c_j c_k,\\
d_3 &=& c_0 \sum_{j=2}^{n-4} \sum_{k=j+2}^{n-2} \sum_{l = k+2}^{n} c_j c_k c_l,\nonumber
\end{eqnarray}
where each $d_i$ contains $i$ internal summations.    By generalising to arbitrary order, we can express the coefficients in the numerator of the rational function
in the form
\begin{equation}
d_i = c_0 \sum_{j = 2}^{\tilde{n}} \sum_{k = j+2}^{\tilde{n}+2} \sum_{l = k+2}^{\tilde{n}+4} ... \sum_{r=q+2}^{n-2} \sum_{s=r+2}^{n} c_j c_k c_l ... c_r c_s,
\end{equation}
where $\tilde{n} = n - 2(i-1)$.  Each outer summation begins with a lower limit of $j=2$ and an upper limit of $n-2(i-1)$, while each
inner summation has a lower limit of $s=r+2$ and an upper limit of $n$, where the summation index $r$ has an initial value of $r=2i$.

Similarly, by investigation, the coefficients in the denominator of Equation~(\ref{eq:confrac}) follow a similar pattern where
\begin{eqnarray}
f_0 &=& 1,\nonumber\\
f_1 &=&  \sum_{j=1}^{N+M} c_j =  \sum_{j=1}^n c_j,\nonumber\\
f_2 &=&  \sum_{j=1}^{n-2} \sum_{k=j+2}^{n} c_j c_k,\\
f_3 &=&  \sum_{j=1}^{n-4} \sum_{k=j+2}^{n-2} \sum_{l = k+2}^{n} c_j c_k c_l.\nonumber\\
\end{eqnarray}
where for each $f_i$, there are $i$ internal summations.  Once again, generalising to arbitrary order, we can 
write the coefficients of the denominator in the form
\begin{equation}
f_i =  \sum_{j = 1}^{\tilde{n}} \sum_{k = j+2}^{\tilde{n}+2} \sum_{l = k+2}^{\tilde{n}+4} ... \sum_{r=q+2}^{n-2} \sum_{s=r+2}^{n} c_j c_k c_l ... c_r c_s.
\label{eq:fi}
\end{equation}
where again $\tilde{n} = n - 2(i-1)$.  This time the lower limit of each external summation begins at unity, with an upper limit again
of $n-2(i-1)$, while the internal summation again has lower and upper limits of $s=r+2$ and $n$ respectively, where in this case, the summation index
$r$ has an initial value of $r=2i-1$.  

\section{Calculating the {\emph {C-fraction}} coefficients.}
Now that we have the coefficients for the rational function, we can equate the original power series with the rational function, i.e.
\begin{equation}
\sum_{i=0}^n a_i x^i=\sum_{i=0}^N d_i x^i\left/ \sum_{i=0}^M f_i x^i\right. .
\end{equation}
By multiplying both sides by the denominator on the right hand side, we define a new equality  
\begin{equation}
\sum_{i=0}^n b_i x^i =\sum_{i=0}^N d_i x^i,
\label{eq:bi}
\end{equation}
where the series on the left hand side is a truncated series of order $n$, i.e.
\begin{equation}
\sum_{i=0}^{n}b_i x^i = \sum_{i=0}^n a_i x^i  \sum_{i=0}^M f_i x^i + \mathcal{O}(x^{n+1}),
\end{equation}
and where we define the coefficients $b_i$ by
\begin{equation}
b_i = a_i + \sum_{j=1}^{y} a_{i-j}f_j.
\label{eq:bi2}
\end{equation}
Here $y=\min(i,\lceil n/2 \rceil)$, where the ceiling $\lceil x \rceil$ of any real number $x$ is the unique integer $m$ satisfying $\lceil x \rceil = m\,\,\,\iff\,\,\,x \leq m < x+1$.   Thus for each order of approximation $n$ we obtain a system of $n$ equations for each $b_i$ of the form $b_i = d_i$.  For $i=0$ we have the trivial solution $b_0 = d_0 \equiv c_0$.  However, at higher orders, there are a number of subtleties that allow us to simplify the above system of equations.

If we look at the case of $b_1$, when $n=1$, we end up with the expression $a_1+a_0 f_1 = a_1+a_0 c_1 = d_1 \equiv 0$, with the simple solution $c_1 = -a_1/a_0$.  Now, if $n=2$ we obtain a similar expression, but in this case $d_1$ is non-zero, i.e.
\begin{equation}
a_1 + a_0 f_1 \equiv a_1 + a_0\sum_{j=1}^2 c_j = c_0\sum_{j=2}^{2} c_j.
\end{equation}
Using the fact that $a_0 = c_0$, this expression again reduces to $a_1 + a_0 c_1 = 0$.  In general, as we go to higher orders of $n$, we end up with an expression of the form
\begin{eqnarray}
b_1 &\equiv& a_1 + a_0\sum_{j=1}^n c_j = c_0\sum_{j=2}^{n} c_j \nonumber \\ &\Rightarrow& a_1 + a_0 c_1 = 0\,\,\,\,\,\,\,\, \forall \,\,\, n \geq 2. 
\end{eqnarray}
As a further example, we look at the coefficient $b_2$.  In this case if $N < i=2$, $d_i = 0$.  Thus at $n=2$ and $n=3$, we end up with the equations
\begin{eqnarray}
&&a_2 + a_1 f_1 = 0, \\
&&a_2 + a_1 f_1 + a_0 f_2 = 0.
\end{eqnarray}
The first equality for $n=2$ provides the solution $a_2 + a_1 (c_1 + c_2) = 0$, while for $n=3$, we obtain
\begin{eqnarray}
&&a_2 + a_1 f_1 + a_0 f_2 \equiv a_2 + a_1\sum_{j=1}^3 c_j + a_0\sum_{j=1}^{1}\sum_{k=3}^{3}c_j c_k = 0,\nonumber\\
&&\Rightarrow a_2+a_1\sum_{j=1}^3 c_j + a_0 c_1 c_3 \equiv a_2 + a_1 (c_1 + c_2) = 0,
\end{eqnarray}
as $a_0 c_1 = -a_1$.  At $n=4$, we obtain our first non-zero value of $d_2$ giving the equation $a_2 + a_1 f_1 + a_0 f_2 = d_2$.  Expressing both sides in terms of $a_i$'s and $c_i$'s we get
\begin{eqnarray}
a_2+ a_1\sum_{j=1}^4 c_j + a_0\sum_{j=1}^{2}\sum_{k=j+2}^{4}c_j c_k = c_0 \sum_{j=2}^{2}\sum_{k=j+2}^{4}c_j c_k.
\end{eqnarray}
Subtracting the quantity on the right hand side gives us the expression
\begin{eqnarray}
&&a_2 + a_1\sum_{j=1}^4 c_j + a_0\sum_{j=1}^{1}\sum_{k=j+2}^{4}c_j c_k = 0\nonumber\\
&&\Rightarrow a_2 + a_1\sum_{j=1}^4 c_j + a_0 c_1\sum_{k=3}^{4} c_k \equiv a_2 + a_1 (c_1 + c_2) = 0,
\end{eqnarray}
where again we have used the fact that $a_0 c_1 = -a_1$.  Thus, we can see that regardless of the value of $n$, we end up with the same equation for $b_1$.  It is also clear from Equation~(\ref{eq:bi2}) that initially each $b_i$ is a function of the coefficients $c_j,\,\,\,\forall\,\,\, 0 \leq j \leq n$ through the functions $f_i$.  However, as we have demonstrated in the examples shown above (and others not displayed here), cancellations between the $c_j$'s manage to kill the higher order coefficients at $i < j \leq n$, meaning that finally, the $b_i$ coefficients are a function of the coefficients $c_j,\,\,\,\forall\,\,\, 0 \leq j \leq i$.  Thus, we can now write the system of equations at each order in the form
\begin{equation}
\bar{b}_i \equiv a_i + \sum_{j=1}^{\lceil i/2 \rceil} a_{i-j}\bar{f}_j = 0,
\label{eq:bi3}
\end{equation}
where
\begin{equation}
\bar{f}_j =  \sum_{k = 1}^{\bar{i}} \sum_{l = k+2}^{\bar{i}+2} \sum_{m = l+2}^{\bar{i}+4} ... \sum_{r=q+2}^{i-2} \sum_{s=r+2}^{i} c_k c_l c_m ... c_r c_s,
\end{equation}
and $\bar{i}=i-2(j-1)$.  As this system of equations are linear in $c_i$, we can express Equation~(\ref{eq:bi3}) in the form
\begin{equation}
\alpha_i + \beta_i c_i = 0.
\end{equation}
By factoring out the $c_i$ coefficient in Equation~(\ref{eq:bi2}) and by equating this expression to zero, we can define a system of iterative linear equations
\begin{equation}
a_i + \sum_{j=1}^{\lceil i/2 \rceil} a_{i-j}\bar{f}_j \equiv a_i + \sum_{j=1}^{\lfloor i/2 \rfloor} a_{i-j}\tilde{f}_j + c_i \left(a_{(i-1)} + \sum_{j=1}^{\lfloor (i-1)/2 \rfloor} a_{(i-1)-j}\hat{f}_j  \right)=0\,\,\,\,\,\,\, \forall\,\,\ i \geq 1.
\end{equation}
Here the floor $\lfloor x \rfloor$ of any real number $x$ is the unique integer satisfying $\lfloor x \rfloor = m\,\,\,\iff\,\,\,x-1 < m \leq x$.  In the above expression we define the reduced coefficients $\bar{f}_j$ as
\begin{equation}
\tilde{f}_j =  \sum_{k = 1}^{\tilde{i}} \sum_{l = k+2}^{\tilde{i}+2} \sum_{m = l+2}^{\tilde{i}+4} ... \sum_{r=q+2}^{(i-1)-2} \sum_{s=r+2}^{i-1} c_k c_l c_m ... c_r c_s
\end{equation}
where $\tilde{i} = (i-1)-2(j-1)$.  While these coefficients have the same form as the $f_i$ coefficients given in Equation~(\ref{eq:fi}), we should note that the upper limit is now a function of $(i-1)$ instead of $n$.  As an example, the $\tilde{f}_i$ coefficients to order $n=3$ are given by
\begin{equation}
\tilde{f}_0 = 1\,\,\, , \,\,\,\tilde{f}_1 = \sum_{k=1}^{i-1} c_k\,\,\, , \,\,\,\tilde{f}_2 = \sum_{k=1}^{(i-1)-2}\sum_{l=k+2}^{(i-1)} c_k c_l\,\,\, ,\,\,\, \tilde{f}_3 = \sum_{k=1}^{(i-1)-4}\sum_{l=k+2}^{(i-1)-2} \sum_{m=l+2}^{(i-1)} c_k c_l c_m\,\,\, , \,\,\,etc.
\end{equation}
We also define the double-reduced coefficients $\hat{f}_j$, with again the same form as the coefficients in Equation~(\ref{eq:fi}), except this time the maximum upper limit is given by $(i-2)$ instead of $n$, i.e.  
\begin{equation}
\hat{f}_j =  \sum_{k = 1}^{\hat{i}} \sum_{l = k+2}^{\hat{i}+2} \sum_{m = l+2}^{\hat{i}+4} ... \sum_{r=q+2}^{(i-2)-2} \sum_{s=r+2}^{i-2} c_k c_l c_l ... c_r c_s
\end{equation}
where $\hat{i} = (i-2)-2(j-1)$. We can see that the functions $\hat{f}_j$ has the same form as the functions $\tilde{f}_j$ where we replace $(i-1)$ by $(i-2)$. Again, to $n=3$ order, the coefficients are described by 
\begin{equation}
\hat{f}_0 = 1\,\,\, , \,\,\,\hat{f}_1 = \sum_{k=1}^{i-2} c_k\,\,\, , \,\,\,\hat{f}_2 = \sum_{k=1}^{(i-2)-2}\sum_{l=k+2}^{(i-2)} c_k c_l\,\,\, ,\,\,\, \hat{f}_3 = \sum_{k=1}^{(i-2)-4}\sum_{l=k+2}^{(i-2)-2} \sum_{m=l+2}^{(i-2)} c_k c_l c_m\,\,\, , \,\,\,etc.
\end{equation}
Finally, solving for $c_i$, we can define the expression
\begin{equation}
c_i = -\frac{\alpha_i}{\beta_i} = - \frac{a_i + \displaystyle \sum_{j=1}^{\lfloor i/2 \rfloor} a_{i-j}\bar{f}_j}{a_{(i-1)} + \displaystyle \sum_{j=1}^{\lfloor (i-1)/2 \rfloor} a_{(i-1)-j}\hat{f}_j}.
\label{eq:ci}
\end{equation}
We can see from the above expression, that the numerator and denominator of the $c_i$ coefficient have a similar form, but with the index $i$ in the numerator, and $(i-1)$ in the denominator.  It is not difficult to then show that the denominator of the $c_i$ coefficient is identical to the numerator in the $c_{i-1}$ coefficient.  Therefore, at the $i^{th}$ order we only need to calculate the numerator at that order, i.e.
\begin{equation}
\alpha_i = a_i + \displaystyle \sum_{j=1}^{\lfloor i/2 \rfloor} a_{i-j}\bar{f}_j,
\end{equation}
and can thus define
\begin{equation}
c_i = - \frac{\alpha_i}{\alpha_{i-1}}.
\label{eq:cifast}
\end{equation}
as the coefficients of a {\emph {C-fraction}} representation of a sub-diagonal Pad\'e approximation at arbitrary order.  We have tested the accuracy of the method against value
obtained from the analytic expressions and have found perfect agreement.

\section{Conclusion}
In this work we have looked at a way of calculating the coefficients of a {\emph {C-fraction}} representation of a sub-diagonal Pad\'e approximation at arbitrary order.  Pad\'e 
approximation has become a common method in improving the convergence of a post-Newtonian series for binary compact objects.  A sub-set of the Pad\'e approximant, called a sub-diagonal Pad\'e, is easily expressed as a regular continued fraction, called a {\emph {C-fraction}}.  The advantage of the {\emph {C-fraction}} representation is that, as we expand the generating power series to higher orders of approximation, we only have one new coefficient to calculate.  While it is possible to analytically derive
expressions for the coefficients of the {\emph {C-fraction}}, these expressions become very lengthy even at low orders of approximation.  Numerical methods exist that are based on calculating the ratios of Hankel determinants.  However, if the determinants are ill-conditioned, it makes the approach unstable.  We present here an iterative numerical method that, given the coefficients of the generating power series, allows one to calculate the {\emph {C-fraction}} coefficients at arbitrary order


\begin{thebibliography}{}

\bibitem{ligo} http://www.ligo.caltech.edu.
\bibitem{virgo} http://www.virgo.infn.it.
\bibitem{lcgt} http://www.icrr.u-tokyo.ac.jp/gr/LCGT.html.
\bibitem{ET} http://www.et-gw.eu.
\bibitem{eLISA} http://sci.esa.int.
\bibitem{DIS} T.~Damour, B.~Iyer \& B.~S.~Sathyaprakash, \prd {\bf 57}, 885 (1998).
\bibitem{PS} E.~K.~Porter \& B.~S.~Sathyaprakash, \prd {\bf 71}, 024017 (2005).
\bibitem{Buonanno} A.~Buonanno \& T.~Damour, \prd {\bf 59}, 084006 (1999).
\bibitem{Acton} F.~S.~Acton, {\emph {Mathematical Methods That Work}}, Mathematical Association of America, Washington (1990).
\bibitem{Wall} K.~S.~Wall, {\emph {Analytic Theory of Continued Fractions}}, Von Nostand, New York (1948).
\bibitem{LW} L.~Lorentzen \& H.~Waadeland, {\emph{Continued Fractions with Applications}}, North-Holland (1992)
\bibitem{JT} W.~Jones \& W.~Thron, {\emph{Continued Fractions : Analytic Theory and Applications}} in \emph{Encylopedia of Mathematics and its Applications}, {\bf 4} (1980)



\end{thebibliography}
\end{document}